\documentclass[aps,prd,amssymb,nofootinbib,twocolumn,epsf,floatfix,
               superscriptaddress]{revtex4}
\usepackage[usenames]{color}
\usepackage[normalem]{ulem} 
\usepackage{graphicx} 
\usepackage{subfigure} 
\usepackage{amssymb}  
\usepackage{amsmath}
\usepackage{mathrsfs} 
\usepackage{epstopdf}

\makeatletter
\renewcommand{\p@subfigure}{\thefigure}
\makeatother

\begin{document}
 
%%%%%%%%%%%%%%%%%%%%%%%%%%%%%%%%%%%%%%%%%%%%%%%%%%%%%%%%%%%%%%%%%%%%%%%%%%%%%%%
% Representing Equations of State With Strong
% Hadron-Quark Phase Transitions
%%%%%%%%%%%%%%%%%%%%%%%%%%%%%%%%%%%%%%%%%%%%%%%%%%%%%%%%%%%%%%%%%%%%%%%%%%%%%%%

\title{Representing Equations of State With Strong First-Order
  Phase Transitions}

\author{Lee Lindblom}
\affiliation{Department of Physics, University of California at San
  Diego, San Diego, California, 92093 USA}

\author{Steve M. Lewis}
\noaffiliation

\author{Fridolin Weber}
\affiliation{Department of Physics, University of California at San
  Diego, San Diego, California, 92093 USA}
\affiliation{Department of Physics, San Diego State University,
  San Diego, California, 92182 USA}

\date{\today}
 
\begin{abstract}
Parametric representations of the high-density nuclear equation of
state are used in constructing models for interpreting the
astrophysical observations of neutron stars.  This study explores how
accurately equations of state with strong first-order phase
transitions can be represented using spectral or piecewise analytic
methods that assume no {\it{a priori}} knowledge of the location or the
strength of the phase transition.  The model equations of state used
in this study have phase transitions strong enough to induce a
gravitational instability that terminates the sequence of stable
neutron stars.  These equations of state also admit a second sequence
of stable stars with core matter that has undergone this strong
first-order phase transition (possibly driven by quark deconfinement).
These results indicate that spectral representations generally achieve
somewhat higher accuracy than piecewise analytic representations
having the same number of parameters.  Both types of representation
show power-law convergence at approximately the same rate.
\end{abstract} 

\maketitle

%%%%%%%%%%%%%%%%%%%%%%%%%%%%%%%%%%%%%%%%%%%%%%%%%%%%%%%%%%%%%%%%%%%%%%%%%%%%%%%
% Introduction
%%%%%%%%%%%%%%%%%%%%%%%%%%%%%%%%%%%%%%%%%%%%%%%%%%%%%%%%%%%%%%%%%%%%%%%%%%%%%%%
\section{Introduction}
\label{s:Introduction}

Parametric representations of the nuclear equation of state are used
in the construction of models designed to describe, and to understand
at a deeper level, the astrophysical observations of neutron stars.
For example, these representations can be used to solve the
relativistic inverse stellar structure problem in which the equation
of state of the stellar matter is determined from a knowledge of the
star's macroscopic properties (e.g. their masses and
radii)~\cite{Lindblom2012, Lindblom2014, Lindblom2014a,
  Lindblom2018a}. This inverse structure problem can be solved by
adjusting the equation of state parameters to minimize the differences
between the neutron-star observations and the models of those observed
properties constructed from the equation of state representation.

The quantity and quality of the astrophysical observations of neutron
stars has improved significantly over the past decade.  The masses and
radii of neutron stars have now been measured at the few percent level
for dozens of neutron stars from observations of binary systems.  Both
mass and radius have been measured for a few individual neutron stars
at the 10-20\% level using x-ray observations~\cite{Ozel2016a,
  Riley2019, Miller2019, Miller2021, Biswas2021, Vinciguerra2024,
  Salmi2024, Choudhury2024, Chatziioannou2024}.  And both mass and
tidal deformability have also been measured using gravitational wave
observations of binary merger events~\cite{De2018, De2018a,
  Chatziioannou2020}, presently at somewhat lower accuracy than that
achieved by the best x-ray observations.

Reliable parametric representations of the high density neutron-star
equation of state are needed to construct useful models of these
observations.  These representations must be capable of representing
the large class of possible neutron-star equations of state at an
accuracy level sufficient for the current observations.  And these
representations should be flexible enough to accommodate more accurate
future observations as they become available.  A number of such
representations have been proposed, for example see
Refs.~\cite{Read:2008iy, Lindblom2010, Lindblom2018, Annala2020,
  Lindblom2022, Lindblom2024b}.  Some of these use spectral methods to
represent the equation of state over the full range of pressures
expected in neutron-star cores, while others use piecewise analytic
representations on a collection of shorter pressure intervals that
together span the range of core pressures.  Both types of
representations have been shown to be convergent (in the sense that
their accuracies can be increased simply by increasing the number of
parameters) for a large collection of nuclear-theory based equations
of state~\cite{Lindblom2018}, including examples with a wide range of
phase transitions~\cite{Lindblom2024, Lindblom2024b}.  These
representations have also been shown to be accurate enough to allow
solutions of the relativistic inverse structure problem with
accuracies commensurate with the accuracy of the available data from
neutron-star observations~\cite{Lindblom2025}.

Spectral methods provide the most efficient representations of smooth
equations of state, converging exponentially as the number $N$ of
spectral basis functions is increased (i.e. with errors decreasing
faster than any power of $1/N$).  In comparison, piecewise analytic
representations typically have power-law convergence (i.e. with errors
decreasing as $1/N^k$ for some particular $k$).  It is less widely
appreciated that spectral representations of non-smooth functions
(e.g. equations of state with first- or second-order phase
transitions) are still convergent, but with power-law rather than
exponential convergence rates~\cite{Boyd1999}. 

Tests of representations of neutron-star equations of state with phase
transitions of various sizes~\cite{Lindblom2024, Lindblom2024b} showed
that even for these cases, the spectral representations were generally
more accurate than the piecewise analytic representations having the
same number of adjustable parameters.  The sizes of the first-order
phase transitions included in those studies were limited to
discontinuities small enough that they could occur within the
contiguous family of neutron stars.  Sufficiently strong first-order
phase transitions trigger an instability that terminates the sequence
of stable neutron stars~\cite{Lindblom98a}.  These instability
triggering first-order phase transitions were also studied in
Refs.~\cite{Lindblom2024, Lindblom2024b}, but representations of
equations of state with stronger first-order phase transitions have
not yet been studied.

There has been some interest recently in the possibility that strong
phase transitions could in some cases lead to a disconnected sequence
of stable higher density relativistic stars~\cite{Paschalidis2018,
  Essick2020, Annala2020}.  The accuracy and convergence properties of
spectral representations of such equations of state have not been
determined at this time.  While the efficacy of using spectral
representations of neutron-star equations of state with phase
transitions has been demonstrated in Refs.~\cite{Lindblom2024,
  Lindblom2024b}, those studies have been strongly criticized (by a
subset of the relativistic astrophysics community) because they did
not include phase transitions strong enough to trigger the
gravitational instability while allowing a disconnected sequence of
stable higher density stellar models.  The primary motivation for the
present study is to determine whether that criticism is justified: Are
spectral methods useful for representing neutron star equations of
state with strong hadron-quark type phase transitions?

The model equations of state used in this study include phase
transitions intended to describe the possible transition from hadron
to quark matter~\cite{Paschalidis2018}.  These phase transitions are
strong enough to terminate the contiguous sequence of neutron stars,
and they allow a second disconnected sequence of stable relativistic
stars with quark matter cores.  Section~\ref{s:ModelEquationsOfState}
describes in detail the model equations of state used in this study.
These equations of state are based on the ACB4 and ACB5 models of
hadron-quark phase transitions introduced in
Ref.~\cite{Paschalidis2018}.  The versions of these equations of state
used in this study are those adapted in Ref.~\cite{Blaschke2020} using
a multi-polytrope representation with a Maxwell construction.  This
study also considers the generalizations introduced in
Ref.~\cite{Blaschke2020}, which incorporate a mixed-phase region at
the hadron-quark interface, mimicking finite-size effects associated
with a ``pasta'' phase~\cite{Glendenning1992, Glendenning2001}.

The equation of state representations used in this study are a causal
Chebyshev polynomial based spectral
representation~\cite{Lindblom2024b}, and a causal piecewise analytic
representation~\cite{Lindblom2018}.  These representations are
described in Secs.~\ref{s:SpectralRepresentations} and
\ref{s:PiecewiseAnalyticRepresentations} respectively.  They are used
to construct model equations of state that cover the full range of
pressures that exist in the cores of both neutron and quark stars.
Section~\ref{s:TwoZoneRepresentations} introduces two-zone equation of
state representations. These composite representations split the
pressure range at the phase transition point with separate spectral
(or piecewise analytic) representations below and above this point.
These above and below representations are then combined to form a
composite representation which includes three additional parameters
that determine the location and size of the phase transition point.
Two-zone representations of this type have not been used previously in
the context of neutron star equations of state.  Determining whether
these two-zone representations provide a more efficient way to
represent equations of state with strong first-order phase transitions
is another primary motivation for this study.

Section~\ref{s:OptimalRepresentations} of this study describes how the
optimal representations of the hadron-quark equations of state are
constructed for this study. Section~\ref{s:Numerical Results}
describes the numerical accuracy and convergence results obtained for
the model equations of state with hadron-quark phase transitions
described in Sec.~\ref{s:ModelEquationsOfState}, using the various
representation methods described in
Sec.~\ref{s:CausalRepresentations}.  These results are discussed in
Sec.~\ref{s:Discussion}.  They show that the accuracy and convergence
rates of the spectral and the piecewise analytic representations are
comparable, with the spectral representations generally being somewhat
more accurate than the piecewise analytic representations having the
same number of adjustable parameters.  The two zone representations
are expected to become more accurate than their one zone counterparts
in the limit of large parameter numbers.  However the tests performed
in this study using eleven or fewer parameters show that the one zone
spectral and piecewise analytic representations are more accurate than
their two zone counterparts with the same number of parameters.

%%%%%%%%%%%%%%%%%%%%%%%%%%%%%%%%%%%%%%%%%%%%%%%%%%%%%%%%%%%%%%%%%%%%%%%%%%%%%%%
%%%%%%%%%%%%%%%%%%%%%%%%%%%%%%%%%%%%%%%%%%%%%%%%%%%%%%%%%%%%%%%%%%%%%%%%%%%%%%%
% Model Equations of State
%%%%%%%%%%%%%%%%%%%%%%%%%%%%%%%%%%%%%%%%%%%%%%%%%%%%%%%%%%%%%%%%%%%%%%%%%%%%%%%
\section{Model Equations of State}
\label{s:ModelEquationsOfState}

The model equations of state used in this study describe possible
strong hadron-quark phase transitions in neutron-star matter.  This
study uses the ACB4 and ACB5 equations of state introduced in
Ref.~\cite{Paschalidis2018}. These equations of state feature
first-order phase transitions, which induce gravitational
instabilities leading to the termination of the stable hadronic
neutron star sequence at approximately $2.0 M_\odot$ for ACB4 and $1.4
M_\odot$ for ACB5. These equations of state also admit a disconnected
branch of stable hybrid stars whose cores consist of the higher
density quark matter.  The ACB4$\,({\Delta_\mathrm{P}})$ and
ACB5$\,({\Delta_\mathrm{P}})$ equations of state introduced in
Ref.~\cite{Blaschke2020} are generalizations of ACB4 and ACB5 that use
second-order phase transitions to model a mixed phase region where the
transition from hadron to quark matter takes place over a range of
pressures determined by the parameter $\Delta_\mathrm{P}$.  These
smoother transitions are designed to mimic ``pasta'' structures in
which hadron and quark matter coexist at the same pressure, see
Refs.~\cite{Glendenning1992, Glendenning2001, Abgaryan2018}.  The
parameter value $\Delta_\mathrm{P}=0$ corresponds to the original ACB4
and ACB5 equations of state with first-order phase transitions, while
the larger $\Delta_\mathrm{P}$ values describe smoother more gradual
second-order transitions between the hadron and quark phases.

\begin{figure}[!t]
    \includegraphics[width=0.4\textwidth]
%                    {../EquationsOfState/ACB4_EOS_Range.eps}
                    {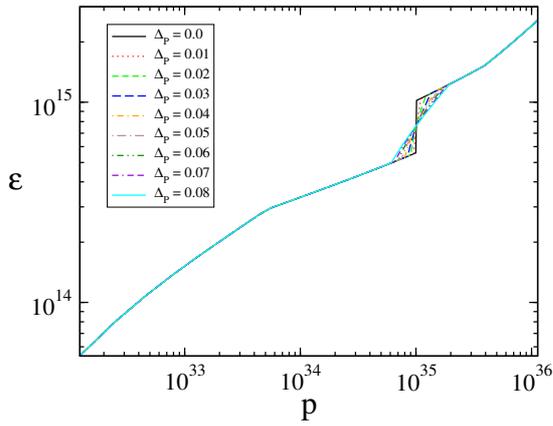}
          \caption{This figure illustrates the
            ACB4$(\Delta_\mathrm{P})$ equations of state that have
            strong hadron-quark phase transitions at the central
            pressure of a 2.0 $M_\odot$ neutron star model.  The total
            energy density $\epsilon$ and pressure $p$ are expressed
            in cgs units: g/cm${}^3$ and erg/cm${}^3$ respectively.
            The parameter $\Delta_\mathrm{P}$ determines the width of
            the pressure region over which the phase transitions
            occur.
          \label{f:ACB4EOSRange}}
\end{figure}

\begin{figure}[h]
    \includegraphics[width=0.4\textwidth]
%          {../EquationsOfState/ACB5_EOS_Range.eps}
                    {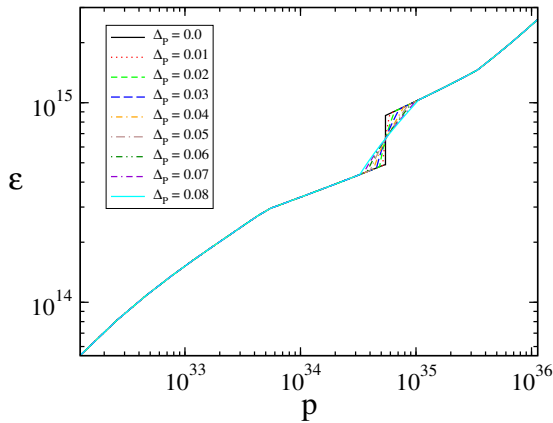}
          \caption{This figure illustrates the
            ACB5$(\Delta_\mathrm{P})$ equations of state that have
            strong hadron-quark phase transitions at the central
            pressure of a 1.4 $M_\odot$ neutron star model.  The
              total energy density $\epsilon$ and pressure $p$ are
              expressed in cgs units: g/cm${}^3$ and erg/cm${}^3$
              respectively.  The parameter $\Delta_\mathrm{P}$
            determines the width of the pressure region over which the
            phase transitions occur.
          \label{f:ACB5EOSRange}}
\end{figure}

\begin{figure}[!t]
    \includegraphics[width=0.4\textwidth]
%          {../EquationsOfState/ACB4_EOS_Detail.eps}
                    {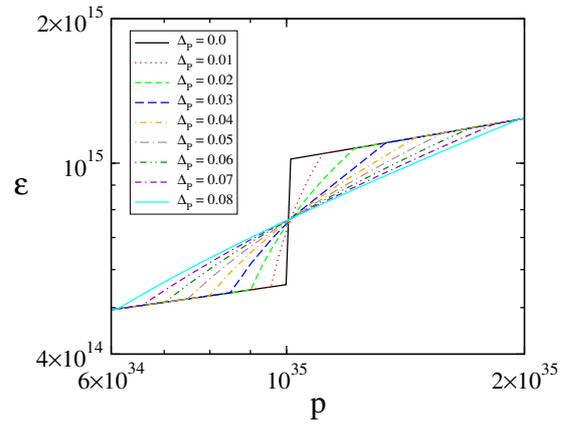}
          \caption{This figure illustrates in more detail the regions
            of the ACB4$(\Delta_\mathrm{P})$ equations of state where the
            strong hadron-quark phase transitions occur.  The
              total energy density $\epsilon$ and pressure $p$ are
              expressed in cgs units: g/cm${}^3$ and erg/cm${}^3$
              respectively.  The parameter $\Delta_\mathrm{P}$
            determines the width of the pressure region over which the
            phase transitions occur.
          \label{f:ACB4EOSDetail}}
\end{figure}

\begin{figure}[!t]
    \includegraphics[width=0.4\textwidth]
%          {../EquationsOfState/ACB5_EOS_Detail.eps}
                    {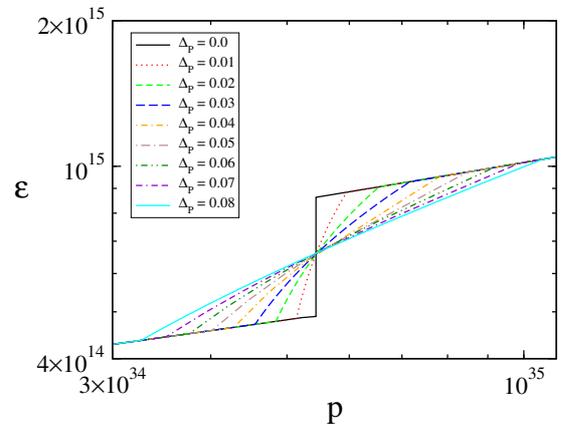}
          \caption{This figure illustrates in more detail the regions
            of the ACB5$(\Delta_\mathrm{P})$ equations of state where
            the strong hadron-quark phase transitions occur.  The
            total energy density $\epsilon$ and pressure $p$ are
            expressed in cgs units: g/cm${}^3$ and erg/cm${}^3$
            respectively.  The parameter $\Delta_\mathrm{P}$
            determines the width of the pressure region over which the
            phase transitions occur.
            \label{f:ACB5EOSDetail}}          
\end{figure}

Figures~\ref{f:ACB4EOSRange} and \ref{f:ACB5EOSRange} illustrate the
ACB4$(\Delta_\mathrm{P})$ and ACB5$(\Delta_\mathrm{P})$ equations of
state over the range of pressures used in this study to evaluate the
accuracy of spectral and piecewise analytic representations.  The
lower end of this range, $p_\mathrm{min}=1.23592430\times 10^{32}$
erg/cm${}^3$, was chosen to correspond to the density,
$\epsilon_\mathrm{min}=5.41165156\times 10^{13}$ g/cm${}^3$, which is
about one fifth normal nuclear density.  The upper end of the pressure
range, $p_\mathrm{max}=1.13139676\times 10^{36}$ erg/cm${}^3$,
corresponds to the central pressures of the maximum mass quark stars
that can be constructed from the ACB4$(\Delta_\mathrm{P})$ equations
of state.  Figures~\ref{f:ACB4EOSDetail} and \ref{f:ACB5EOSDetail}
illustrate in more detail the regions of the ACB4$(\Delta_\mathrm{P})$
and ACB5$(\Delta_\mathrm{P})$ equations of state where the phase
transitions occur.
\vfill\break

\begin{figure}[!t]
    \includegraphics[width=0.32\textwidth]
%          {../MR_Curves/ACB4_MR.eps}
                    {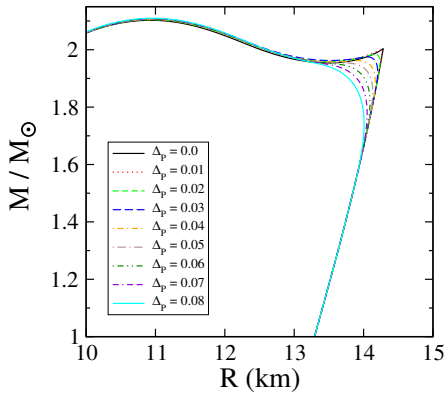}
          \caption{This figure illustrates the mass-radius curves
            obtained by solving the Oppenheimer-Volkoff relativistic
            stellar structure equations using the ACB4$(\Delta_\mathrm{P})$
            equations of state.
          \label{f:ACB4MRCurves}}
\end{figure}

\begin{figure}[!t]
    \includegraphics[width=0.32\textwidth]
%          {../MR_Curves/ACB5_MR.eps}
                    {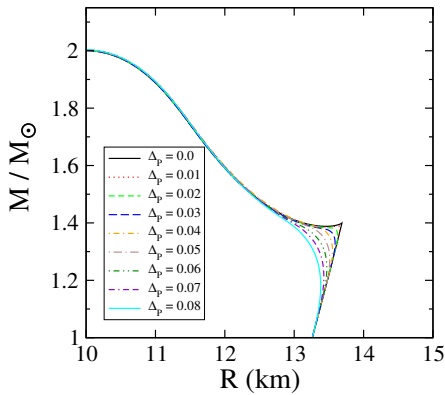}
          \caption{This figure illustrates the mass-radius curves
            obtained by solving the Oppenheimer-Volkoff relativistic
            stellar structure equations using the ACB5$(\Delta_\mathrm{P})$
            equations of state.
          \label{f:ACB5MRCurves}}
\end{figure}

Figures~\ref{f:ACB4MRCurves} and \ref{f:ACB5MRCurves} illustrate the
mass-radius curves obtained by solving the Oppenheimer-Volkoff
relativistic stellar structure equations~\cite{Oppenheimer1939} using
the ACB4$(\Delta_\mathrm{P})$ and ACB5$(\Delta_\mathrm{P})$ equations
of state respectively.  The stellar models lying between the maxima
and the subsequent minima of some of these mass-radius curves are
unstable to a general relativistic gravitational instability that
triggers collapse to a black hole.  The curves in
Fig.~\ref{f:ACB4MRCurves} show that the ACB4$(\Delta_\mathrm{P})$
equations of state with $\Delta_\mathrm{P} < 0.05$ have maxima that
trigger this instability, while also admitting stable composite
hadron-quark stars with larger central densities.  Similarly the
curves in Fig.~\ref{f:ACB5MRCurves} show that the
ACB5$(\Delta_\mathrm{P})$ equations of state with $\Delta_\mathrm{P} <
0.02$ trigger this instability, while also admitting stable composite
hadron-quark stars with larger central densities.  The maximum masses
and central pressures of the stable composite hadron-quark stars
obtained using either the ACB4$(\Delta_\mathrm{P})$ or the
ACB5$(\Delta_\mathrm{P})$ equations of state are very insensitive to
the width of the phase transition region determined by the parameter
$\Delta_\mathrm{P}$.
%

%%%%%%%%%%%%%%%%%%%%%%%%%%%%%%%%%%%%%%%%%%%%%%%%%%%%%%%%%%%%%%%%%%%%%%%%%%%%%%%
%%%%%%%%%%%%%%%%%%%%%%%%%%%%%%%%%%%%%%%%%%%%%%%%%%%%%%%%%%%%%%%%%%%%%%%%%%%%%%%
% Causal Parametric Representations
%%%%%%%%%%%%%%%%%%%%%%%%%%%%%%%%%%%%%%%%%%%%%%%%%%%%%%%%%%%%%%%%%%%%%%%%%%%%%%%
\section{Causal Parametric Equation of State Representations}
\label{s:CausalRepresentations}
 
This section describes the general methods used to construct the
causal parametric representations of the high-density neutron-star
equations of state used in this study.  These methods are applied in
Secs.~\ref{s:SpectralRepresentations} and
\ref{s:PiecewiseAnalyticRepresentations} to construct the particular
spectral and the piecewise analytic representations used here.
Section~\ref{s:TwoZoneRepresentations} introduces two zone equation of
state representations that split the pressure range at the phase
transition point, with separate spectral (or piecewise analytic)
representations below and above this point.

The speed of sound, $v$, in a relativistic barotropic fluid is
determined by the equation of state:
$v^2=dp/d\epsilon$~\cite{Landau1959}.  These sound speeds satisfy the
thermodynamic stability condition, $v^2\ge 0$, and causality
condition, $v^2\leq c^2$ (where $c$ is the speed of light), if and
only if the velocity function $\Upsilon$,
\begin{equation}
  \Upsilon=\frac{c^2-v^2}{v^2},
    \label{e:Upsilon_Def}
\end{equation}
is non-negative, $\Upsilon\geq 0$.

The velocity function $\Upsilon$ is determined by the equation of
state: $\Upsilon(p)=c^2\,d\epsilon/dp -1$.  Conversely, $\Upsilon(p)$
can be used as a generating function from which the standard equation
of state, $\epsilon=\epsilon(p)$, can be determined by quadrature.
The definition of the velocity function $\Upsilon(p)$ can be
re-written as the ordinary differential equation,
\begin{equation}
  \frac{d\epsilon(p)}{dp} =\frac{1}{c^2}+\frac{\Upsilon(p)}{c^2}.
  \label{e:Upsilon_p_Def}
\end{equation}
which can then be integrated to determine the equation of state,
$\epsilon=\epsilon(p)$:
\begin{equation}
  \epsilon(p)=\epsilon_\mathrm{min}+\frac{p-p_\mathrm{min}}{c^2} 
  +\frac{1}{c^2}\int_{p_\mathrm{min}}^p\!\!\!\Upsilon(p')dp'.
  \label{e:epsilon_p}
\end{equation}
The velocity functions $\Upsilon(p,\upsilon_a)$ used in this study
depend on a series of parameters $\upsilon_a$, for $1\leq a \leq
N_\mathrm{parms}$.  Equation~(\ref{e:epsilon_p}) therefore determines
a family of equations of state, $\epsilon=\epsilon(p,\upsilon_a)$,
whose parameters can be adjusted to model realistic physical equations
of state.

The process of finding the optimal choice of parameters needed to fit
a particular equation of state generally requires a knowledge of how
the parametric equation of state, $\epsilon(p,\upsilon_a),$ changes as
the parameters, $\upsilon_a$, are varied.  In some cases the most
efficient way to determine $\partial\epsilon(p,\upsilon_b)/
\partial\upsilon_a$ is to evaluate the integrals,
\begin{equation}
  \frac{\partial\epsilon(p,\upsilon_b)}{\partial \upsilon_a}=
  \frac{1}{c^2}\int_{p_\mathrm{min}}^p\!\!\!\frac{\partial\Upsilon(p',\upsilon_b)}
       {\partial\upsilon_a}dp'.
  \label{e:depsilon_p}
\end{equation}
In more complicated cases it may be more efficient to evaluate
these derivatives numerically using finite difference
expressions, e.g.,
\begin{equation}
  \frac{\partial\epsilon(p,\upsilon_a)}{\partial \upsilon_a}=
  \frac{\epsilon(p,\upsilon_a+\delta\upsilon_a)
    -\epsilon(p,\upsilon_a-\delta\upsilon_a)}{2\delta\upsilon_a}
  +{\mathcal O}(\delta\upsilon_a^2).
  \label{e:depsilon_p2}
\end{equation}

%%%%%%%%%%%%%%%%%%%%%%%%%%%%%%%%%%%%%%%%%%%%%%%%%%%%%%%%%%%%%%%%%%%%%%%%%%%%%%%
%%%%%%%%%%%%%%%%%%%%%%%%%%%%%%%%%%%%%%%%%%%%%%%%%%%%%%%%%%%%%%%%%%%%%%%%%%%%%%%
% Spectral Representations
%%%%%%%%%%%%%%%%%%%%%%%%%%%%%%%%%%%%%%%%%%%%%%%%%%%%%%%%%%%%%%%%%%%%%%%%%%%%%%%
\subsection{Spectral Representations}
\label{s:SpectralRepresentations}

Causal parametric representations of the neutron-star equation of
state can be constructed by expressing the velocity function,
$\Upsilon(p,\upsilon_a)$, as a spectral expansion based on Chebyshev
polynomials (see Ref.~\cite{Lindblom2024b}):
\begin{equation}
  \Upsilon(p,\upsilon_a)=\Upsilon_\mathrm{min}\exp\left\{
  \sum_{a=0}^{N_\mathrm{parms}-1}\upsilon_a(1+y)T_a(y)\right\},
  \label{e:Upsilon_p_Chebyshev}
\end{equation}
where the $T_a(y)$ are Chebyshev polynomials.  The variable $y$
(defined below) is a function of the pressure having the property that
$y=-1$ when $p=p_\mathrm{min}$. The constant $\Upsilon_\mathrm{min}$
is evaluated from the low-density equation of state at the point
$p=p_\mathrm{min}$ where it matches onto the high density spectral
representation determined by Eq.~(\ref{e:Upsilon_p_Chebyshev}).
Choosing $\Upsilon_\mathrm{min}$ in this way ensures that no
artificial second-order phase-transition discontinuity is introduced
at the matching point.  These expansions guarantee that
$\Upsilon(p)\geq 0$ for every choice of $\upsilon_a$. Therefore any
equation of state determined from one of these
$\Upsilon(p,\upsilon_a)$ automatically satisfies the causality and
thermodynamic stability conditions.

Chebyshev polynomials are defined by the recursion relation
$T_{a+1}(y)=2yT_a(y)-T_{a-1}(y)$ with $T_0(y)=1$ and $T_1(y)=y$.
Spectral expansions using Chebyshev basis functions are well behaved
on the domain $-1\leq y \leq 1$~\cite{Boyd1999}.  The variable $y$
used here has been chosen to be the following linear function of $\log
p$,
\begin{equation}
  y(p) = -1 + 2\log\left(\frac{p}{p_\mathrm{min}}\right)\left[
    \log\left(\frac{p_\mathrm{max}}{p_\mathrm{min}}\right)\right]^{-1},
  \label{e:zDef}
\end{equation}
to ensure that $-1\leq y\leq 1$ for pressures in the range
$p_\mathrm{min}\leq p \leq p_\mathrm{max}$.  The factor $1+y$ that
appears in Eq.~(\ref{e:Upsilon_p_Chebyshev}) ensures that
$\Upsilon(p,\upsilon_a)$ has the limit,
$\Upsilon(p_\mathrm{min},\upsilon_a)=\Upsilon_\mathrm{min}$, for every
choice of spectral parameters $\upsilon_a$.  The partial derivatives
$\partial\epsilon(p,\upsilon_b)/\partial\upsilon_a$ needed to evaluate
the integrals in Eq.~(\ref{e:depsilon_p}) for this spectral
representation are given by:
\begin{equation}
  \frac{\partial\Upsilon(p,\upsilon_b)}{\partial\upsilon_a} = [1+y(p)]
  T_a[y(p)] \Upsilon(p,\upsilon_b).
\end{equation}

%%%%%%%%%%%%%%%%%%%%%%%%%%%%%%%%%%%%%%%%%%%%%%%%%%%%%%%%%%%%%%%%%%%%%%%%%%%%%%%
%%%%%%%%%%%%%%%%%%%%%%%%%%%%%%%%%%%%%%%%%%%%%%%%%%%%%%%%%%%%%%%%%%%%%%%%%%%%%%%
% Piecewise Analytic Representations
%%%%%%%%%%%%%%%%%%%%%%%%%%%%%%%%%%%%%%%%%%%%%%%%%%%%%%%%%%%%%%%%%%%%%%%%%%%%%%%
\subsection{Piecewise Analytic Representations}
\label{s:PiecewiseAnalyticRepresentations}

Causal piecewise-analytical representations are constructed by
subdividing the pressure domain $[p_\mathrm{min},p_\mathrm{max}]$ into
$N$ subdomains with
$p_\mathrm{min}=p_0<p_1<...<p_{\scriptscriptstyle{N-1}}<
p_{\scriptscriptstyle{N}}= p_\mathrm{max}$ (see
Ref.~\cite{Lindblom2018}).  In this study the $p_k$ are chosen to be
distributed logarithmically so that $p_k
=p_{k-1}(p_\mathrm{max}/p_\mathrm{min})^{1/N}$ for each $k=1$, ...,
$N$. In these representations the velocity function $\Upsilon(p)$ is
taken to be a simple power law of the pressure:
\begin{equation}
 \Upsilon(p,\upsilon_k)=
 \Upsilon_k\left(\frac{p}{p_k}\right)^{\upsilon_{k+1}},
\end{equation}
in the subdomain $p_k \leq p \leq p_{k+1}$. The adjustable parameters
$\Upsilon_k$ and $\upsilon_{k+1}$ determine its properties in each
subdomain.  The integral in Eq.~(\ref{e:epsilon_p}) for this
$\Upsilon(p)$ is easy to perform, resulting in the following expression
for $\epsilon(p,\upsilon_k)$
\begin{eqnarray}
  \epsilon(p,\upsilon_k) &=&\epsilon_k+
  \frac{p-p_k}{c^2} \nonumber\\
  &&+ \frac{p_k\Upsilon_k}{(1+\upsilon_{k+1})c^2}
  \left[\left(\frac{p}{p_k}\right)^{1+\upsilon_{k+1}}-1\right]\quad
  \label{e:epsilon_p_a}
\end{eqnarray}
in the pressure subdomain $p_k\leq p < p_{k+1}$.

The constants $\Upsilon_k$ and $\epsilon_k$ for the piecewise analytic
representations studied here are determined iteratively by enforcing
continuity of $\Upsilon(p)$ and $\epsilon(p)$ at the pressure
subdomain boundaries:
\begin{eqnarray}
  \Upsilon_k &=& \Upsilon_{k-1}\left(\frac{p_k}{p_{k-1}}\right)^{\upsilon_{k}},
    \label{e:Upsilon_k}\\
\epsilon_k &=&\epsilon_{k-1} + \frac{p_k-p_{k-1}}{c^2}\nonumber\\
&&+ \frac{p_{k-1}\Upsilon_{k-1}}{(1+\upsilon_{k})c^2}
\left[\left(\frac{p_k}{p_{k-1}}\right)^{1+\upsilon_{k}}-1\right],
\label{e:epsilon_k}
\end{eqnarray}
with $\Upsilon_0=\Upsilon_\mathrm{min}$.  The remaining constants
$\upsilon_k$ for $1\leq k \leq N=N_\mathrm{parms}$ are the independent
parameters that determine the equation of state in each pressure
subdomain.  The analytic expressions for the derivatives
$\partial\epsilon(p,\upsilon_b) /\partial\upsilon_a$ are quite
complicated for these representations, so these derivatives have been
computed numerically for this study using Eq.~(\ref{e:depsilon_p2}).

%%%%%%%%%%%%%%%%%%%%%%%%%%%%%%%%%%%%%%%%%%%%%%%%%%%%%%%%%%%%%%%%%%%%
% Two Zone Representations
%%%%%%%%%%%%%%%%%%%%%%%%%%%%%%%%%%%%%%%%%%%%%%%%%%%%%%%%%%%%%%%%%%%%
\subsection{Two Zone Representations}
\label{s:TwoZoneRepresentations}
%%%%%%%%%%%%%%%%%%%%%%%%%%%%%%%%%%%%%%%%%%%%%%%%%%%%%%%%%%%%%%%%%%%%

The equation of state representations presented in
Secs.~\ref{s:SpectralRepresentations} and
\ref{s:PiecewiseAnalyticRepresentations} provide unified descriptions
of the equation of state on the domain $p_\mathrm{min} \leq p \leq
p_\mathrm{max}$.  When there is a large phase transition at a pressure
$p_\mathrm{pt}$ somewhere in this domain, $p_\mathrm{min} \leq
p_\mathrm{pt} \leq p_\mathrm{max}$, it may be possible to construct
more efficient representations by breaking the original domain into
two separate zones with pressures below and above the phase transition
point.  This can be done by constructing a representation in the first
zone, $p_\mathrm{min} \leq p \leq p_\mathrm{pt}$, as described in
Sec.~\ref{s:SpectralRepresentations} or
\ref{s:PiecewiseAnalyticRepresentations} using the additional
parameter $p_\mathrm{pt}$ that determines the location of the phase
transition.  A separate representation can similarly be constructed in
the second zone, $p_\mathrm{pt} \leq p \leq p_\mathrm{max}$, by
introducing two additional parameters $\epsilon_\mathrm{pt}$ and
$\Upsilon_\mathrm{pt}$ that describe the total energy density and the
sound speed just above the phase transition boundary.  The resulting
two zone representation requires a total of $N_\mathrm{parms}=3+
N^{<}_\mathrm{parms} +N^{>}_\mathrm{parms}$ parameters, where
$N^{>}_\mathrm{parms}$ and $N^{<}_\mathrm{parms}$ are the number of
parameters needed to specify the spectral or piecewise representations
in the zones above and below the phase transition point.

%%%%%%%%%%%%%%%%%%%%%%%%%%%%%%%%%%%%%%%%%%%%%%%%%%%%%%%%%%%%%%%%%%%%
% Constructing Optimal Parametric Representations
%%%%%%%%%%%%%%%%%%%%%%%%%%%%%%%%%%%%%%%%%%%%%%%%%%%%%%%%%%%%%%%%%%%%
\section{Constructing Optimal Parametric Representations}
\label{s:OptimalRepresentations}
%%%%%%%%%%%%%%%%%%%%%%%%%%%%%%%%%%%%%%%%%%%%%%%%%%%%%%%%%%%%%%%%%%%%

Each of the model equations of state used in this study,
ACB4$(\Delta_\mathrm{P})$ and ACB5$(\Delta_\mathrm{P})$, consists of a
table of energy-density pressure pairs: $\{\epsilon_i,p_i\}$ for
$1\leq i \leq N_\mathrm{eos}$.  A parametric representation of one of
these equations of state consists of a function
$\epsilon(p,\upsilon_a)$ and the values of the parameters $\upsilon_a$
for $1\leq a \leq N_\mathrm{parms}$ that determine the energy density
as a function of the pressure.  The accuracy of a particular
representation is determined by evaluating the dimensionless error
residual $\chi(\upsilon_a)$, defined by
\begin{eqnarray}
\chi^2(\upsilon_a)=\sum_{i=1}^{N_\mathrm{eos}}\frac{1}{N_\mathrm{eos}}\left\{
\log\left[\frac{\epsilon(p_i,\upsilon_a)}
  {\epsilon_i}\right]\right\}^2.
\label{e:ResidualDef}
\end{eqnarray}
The optimal choice of the parameters $\upsilon_a$ to represent a
particular equation of state $\{\epsilon_i,p_i\}$ is obtained by
minimizing $\chi(\upsilon_a)$ with respect to variations in each of
the parameters $\upsilon_a$.  This minimization is carried out
numerically in this study using the Levenberg-Marquardt
method~\cite{numrec_f}.  This minimization process requires a
knowledge of the derivatives of $\chi^2(\upsilon_a)$ with respect to
the parameters $\upsilon_a$.  These derivatives are determined by the
derivatives of $\epsilon(p,\upsilon_a)$,
\begin{eqnarray} 
  \frac{\partial\chi^2}{\partial\upsilon_a}=
  \frac{2}{N_\mathrm{eos}}\sum_{i=1}^{N_\mathrm{eos}}
  \frac{\partial\log\epsilon(p_i,\upsilon_b)}
       {\partial\upsilon_a} \log\left[\frac{\epsilon(p_i,\upsilon_b)}
         {\epsilon_i}\right],
\label{e:ResidualDef}
\end{eqnarray}
which in turn are determined for this study using
Eq.~(\ref{e:depsilon_p}) or (\ref{e:depsilon_p2}).

\vspace{0.5cm}

%%%%%%%%%%%%%%%%%%%%%%%%%%%%%%%%%%%%%%%%%%%%%%%%%%%%%%%%%%%%%%%%%%%%
% Numerical Results
%%%%%%%%%%%%%%%%%%%%%%%%%%%%%%%%%%%%%%%%%%%%%%%%%%%%%%%%%%%%%%%%%%%%
\section{Numerical Results}
\label{s:Numerical Results}

Causal spectral representations of the high-density portions of the
ACB4$(\Delta_\mathrm{P})$ and ACB5$(\Delta_\mathrm{P})$ equations of state have been
constructed numerically over a range of pressures, $p_\mathrm{min}\leq
p \leq p_\mathrm{max}$, using the methods described in
Secs.~\ref{s:SpectralRepresentations} and
\ref{s:OptimalRepresentations}.  The lower end of the pressure range,
$p_\mathrm{min}=1.23592430\times 10^{32}$ erg/cm${}^3$, was chosen to
be the point in the exact equation of state tables corresponding to
the density, $\epsilon_\mathrm{min}=5.41165156\times 10^{13}$
g/cm${}^3$, which is about one fifth normal nuclear density.  The
upper limit of this pressure range, $p_\mathrm{max}=1.13139676\times
10^{36}$ erg/cm${}^3$, corresponds to the central pressures of the
maximum mass quark stars that can be constructed from the
ACB4$(\Delta_\mathrm{P})$ equations of state.  The value of the velocity
parameter $\Upsilon_\mathrm{min}=13.7751914$ was chosen to ensure that
the spectral representations do not introduce a non-physical
second-order phase transition at the $p=p_\mathrm{min}$ point.

\begin{figure}[!t]
    \includegraphics[width=0.4\textwidth]
%          {../PFits/ACB4_Chebyshev_Fits.eps}
                    {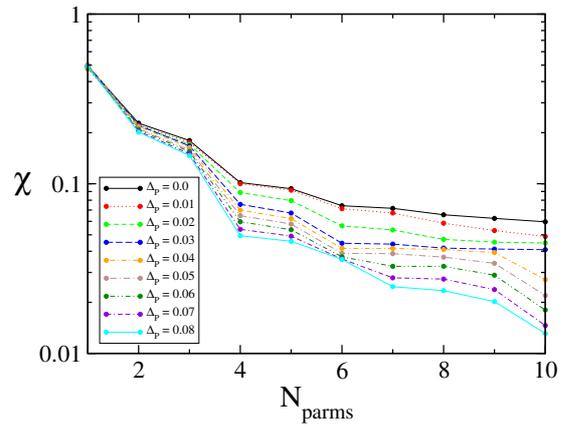}
          \caption{This figure illustrates the equation of state error
            measures, $\chi$, as functions of the number of spectral
            parameters $N_\mathrm{parms}$ for the causal Chebyshev
            polynomial representations of the ACB4$(\Delta_\mathrm{P})$
            equations of state.
          \label{f:ACB4ChebyshevFits}}
\end{figure}

\begin{figure}[!t]
    \includegraphics[width=0.4\textwidth]
%          {../PFits/ACB5_Chebyshev_Fits.eps}
                    {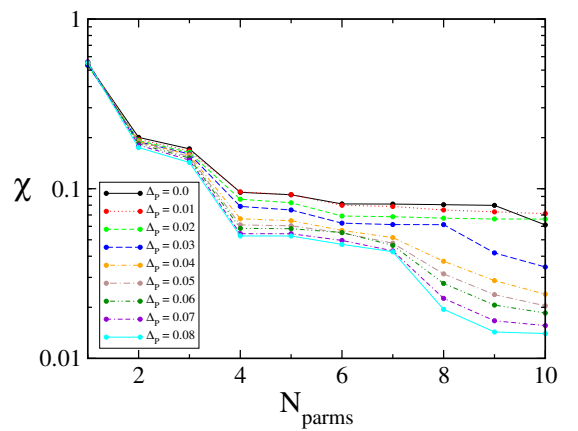}
          \caption{This figure illustrates the equation of state error
            measures, $\chi$, as functions of the number of spectral
            parameters $N_\mathrm{parms}$ for the causal Chebyshev
            polynomial representations of the ACB5$(\Delta_\mathrm{P})$
            equations of state.
          \label{f:ACB5ChebyshevFits}}
\end{figure}

Figures~\ref{f:ACB4ChebyshevFits} and \ref{f:ACB5ChebyshevFits} show
the minimum values of the equation of state error measures, $\chi$
defined in Eq.~(\ref{e:ResidualDef}), as functions of the number of
spectral parameters, $N_\mathrm{parms}$, for the spectral
representations of the ACB4$(\Delta_\mathrm{P})$ and
ACB5$(\Delta_\mathrm{P})$ equations of state.  These results show that
the spectral representations of the equations of state with
second-order phase transitions, i.e. those with $\Delta_\mathrm{P}>0$,
converge more rapidly than the representation of the discontinuous
$\Delta_\mathrm{P}=0$ equation of state.  Nevertheless, the spectral
representations of the $\Delta_\mathrm{P}=0$ equations of state are
convergent with $\chi$ decreasing monotonically from $\chi\approx 0.5$
for $N_\mathrm{parms}=1$ to $\chi\approx 0.06$ for
$N_\mathrm{parms}=10$.

Piecewise analytic representations of the high-density portions of the
ACB4$(\Delta_\mathrm{P})$ and ACB5$(\Delta_\mathrm{P})$ equations of
state have also been constructed numerically using the methods
described in Secs.~\ref{s:PiecewiseAnalyticRepresentations} and
\ref{s:OptimalRepresentations}. The values of the equation of state
parameters used for these representations cover the same pressure
range as those used for the spectral representations:
$p_0=p_\mathrm{min}=1.23592430\times 10^{32}$ erg/cm${}^3$,
$p_{\scriptscriptstyle{N}_\mathrm{parms}}=
p_\mathrm{max}=1.13139676\times 10^{36}$ erg/cm${}^3$,
$\epsilon_0=\epsilon_\mathrm{min}=5.41165156\times 10^{13}$
g/cm${}^3$, and $\Upsilon_0=\Upsilon_\mathrm{min}=13.7751914$.

\begin{figure}[!t]
    \includegraphics[width=0.4\textwidth]
%          {../PFits/ACB4_Piecewise_Analytic_Fits.eps}
                    {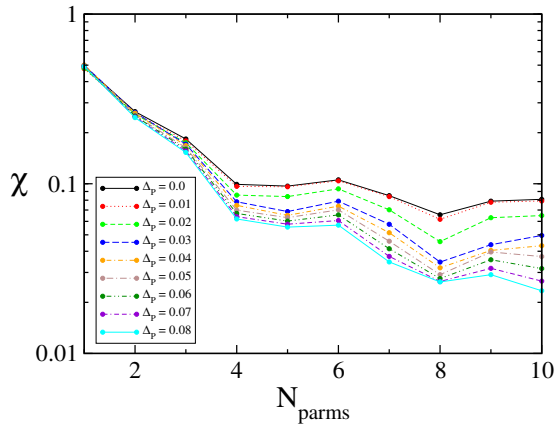}
          \caption{This figure illustrates the equation of state error
            measures, $\chi$, as functions of the number of spectral
            parameters $N_\mathrm{parms}$ for the causal piecewise
            analytic representations of the ACB4$(\Delta_\mathrm{P})$ equations
            of state.
          \label{f:ACB4PiecewiseAnalyticFits}}
\end{figure}

\begin{figure}[!t]
    \includegraphics[width=0.4\textwidth]
%          {../PFits/ACB5_Piecewise_Analytic_Fits.eps}
                    {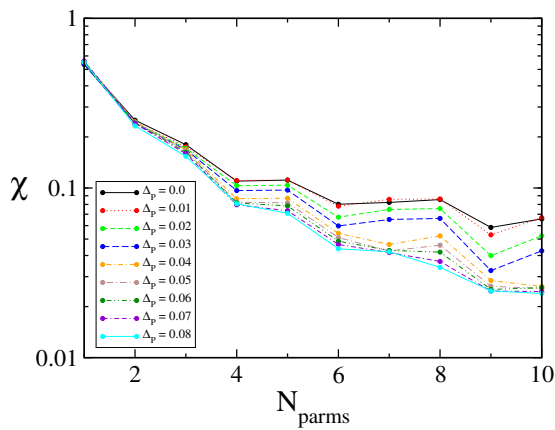}
          \caption{This figure illustrates the equation of state error
            measures, $\chi$, as functions of the number of spectral
            parameters $N_\mathrm{parms}$ for the causal piecewise
            analytic representations of the ACB5$(\Delta_\mathrm{P})$ equations
            of state.
          \label{f:ACB5PiecewiseAnalyticFits}}
\end{figure}

Figures~\ref{f:ACB4PiecewiseAnalyticFits} and
\ref{f:ACB5PiecewiseAnalyticFits} show the minimum values of the
equation of state error measure, $\chi$, defined in
Eq.~(\ref{e:ResidualDef}) as functions of the number of parameters,
$N_\mathrm{parms}$, for the piecewise analytic fits to the
ACB4$(\Delta_\mathrm{P})$ and ACB5$(\Delta_\mathrm{P})$ equations of
state.  Like the spectral representations, these results show that the
piecewise analytic representations of the smoother equations of state
with $\Delta_\mathrm{P}>0$ have smaller errors than than the
representation of the discontinuous $\Delta_\mathrm{P}=0$ equation of
state, and that the values of $\chi$ generally get smaller as
$N_\mathrm{parms}$ increases.  Unlike the results for the spectral
representations, however, the values of $\chi$ for these piecewise
analytic representations are not strictly decreasing as
$N_\mathrm{parms}$ increases.  This non-monotonic behavior appears to
be caused by the changes in the proximity of the closest subdomain
boundary to the phase transition point as $N_\mathrm{parms}$ is
increased.

\begin{figure}[!t]
    \includegraphics[width=0.4\textwidth]
%          {../PFits/Comparison_Fits.eps}
                    {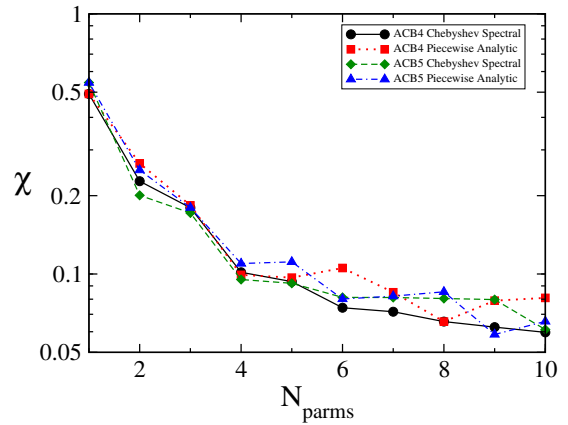}
          \caption{This figure illustrates the error measures
            $\chi$ for the causal Chebyshev spectral and the
            causal piecewise analytic representations of the ACB4 and
            ACB5 equations of state with $\Delta_\mathcal{P}=0$ as functions of
            the number of parameters $N_\mathrm{parms}$.
          \label{f:ComparisonFits}}
\end{figure}

Figure~\ref{f:ComparisonFits} presents a direct comparison between the
causal Chebyshev spectral representations of the discontinuous ACB4
and ACB5 equations of state with $\Delta_\mathrm{P}=0$, and their
causal piecewise analytic representations.  These results show that
the spectral and the piecewise analytic representations have
comparable accuracies.  The values of $\chi$ for the spectral
representations are generally somewhat smaller than those for the
piecewise analytic representations. Therefore the spectral
representations are generally somewhat more accurate than the
piecewise analytic representations with the same number of parameters.

\begin{figure}[!t]
    \includegraphics[width=0.4\textwidth]
%          {../PFits/ACB4_deltaP0_Chebyshev_Diffs.eps}
                    {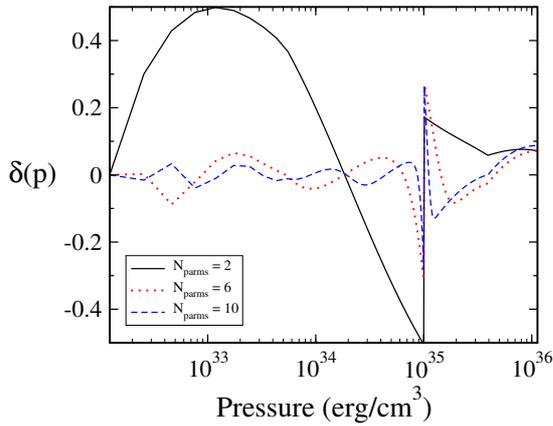}
          \caption{This figure illustrates the fractional errors,
            $\delta(p)=1-
            \epsilon(p,\upsilon_a)/\epsilon_\mathrm{exact}(p)$, for
            the causal Chebyshev spectral representations of the
            ACB4$(\Delta_\mathrm{P}=0)$ equation of state with
            $N_\mathrm{parms}=2$, $6$, or $10$ spectral parameters.
          \label{f:Chebyshev_Diffs}}
\end{figure}

\begin{figure}[!t]
    \includegraphics[width=0.4\textwidth]
%          {../PFits/ACB4_deltaP0_Piecewise_Analytic_Diffs.eps}
                    {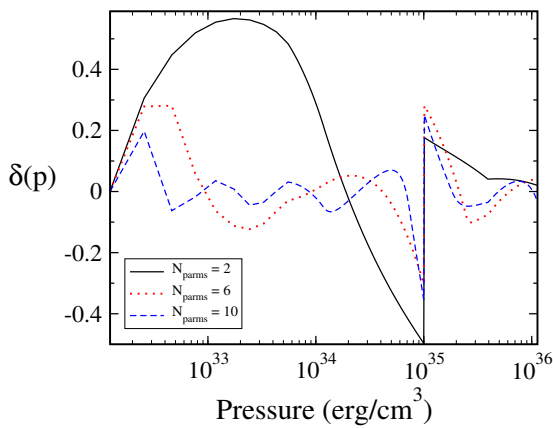}
          \caption{This figure illustrates the fractional errors,
            $\delta(p)=1-
            \epsilon(p,\upsilon_a)/\epsilon_\mathrm{exact}(p)$, for
            the causal piecewise analytic representations of the
            ACB4$(\Delta_\mathrm{P}=0)$ equation of state with
            $N_\mathrm{parms}=2$, $6$, or $10$ parameters.
          \label{f:Piecewise_Analytic_Diffs}}
\end{figure}

Figures~\ref{f:Chebyshev_Diffs} and \ref{f:Piecewise_Analytic_Diffs}
illustrate the pointwise accuracies of the causal Chebyshev spectral
and the causal piecewise analytic representations of the discontinuous
ACB4 equation of state with $\Delta_\mathrm{P}=0$.  These figures show
the pointwise fractional errors, $\delta(p)$, defined by,
\begin{equation}
  \delta(p) = 1 - \frac{\epsilon(p,\upsilon_a)}{\epsilon_\mathrm{exact}(p)}.
\end{equation}
For clarity in these figures, only the errors for the
$N_\mathrm{parms}=2$, 6, and 10 representations are shown.  These
figures illustrate that both the spectral and piecewise analytic
representations make significant errors at and near the phase
transition points, but that both types of representation become more
accurate overall as the number of parameters is increased. The
non-smooth features in these figures near the lower end of the
pressure range are caused by the sparseness of the points in the exact
ACB4 equation of state tables $\{\epsilon_i,p_i\}$ in that region.

\begin{figure}[!t]
    \includegraphics[width=0.4\textwidth]
%          {../TwoZoneFits/TwoZoneConvergence.eps}
                    {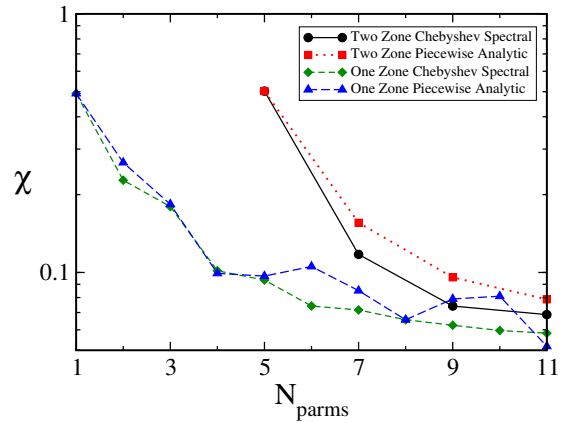}
          \caption{This figure illustrates the equation of state
            accuracy measure $\chi$ as a function of
            $N_\mathrm{parms}$ for one and two zone representations of
            the ACB4$(\Delta_\mathrm{P}=0)$ equation of state based on
            Chebyshev spectral or piecewise analytic methods.  These
            results show that the one zone representations are more
            efficient than the two zone representations for
            $N_\mathrm{parms} \leq 11$.
          \label{f:TwoZoneFits}}
\end{figure}

The two zone spectral and piecewise analytic representations as
described in Secs.~\ref{s:TwoZoneRepresentations} and
\ref{s:OptimalRepresentations} have also been constructed in this
study for the discontinuous ACB4 equation of state with
$\Delta_\mathrm{P}=0$.  The number of parameters in these two zone
representations are the sum of the parameters needed in the two
independent zones plus three additional parameters needed to describe
the location and size of the phase transition point.  Thus the total
number of parameters is given by
$N_\mathrm{parms}=3+N^{<}_\mathrm{parms} +N^{>}_\mathrm{parms}$, where
$N^{>}_\mathrm{parms}$ and $N^{<}_\mathrm{parms}$ are the number of
parameters needed to specify the spectral or piecewise analytic
representations in the regions above and below the phase transition
point.  The values of the three additional parameters needed to
specify the phase transition properties in this study have been set to
the optimal values from the exact ACB4 equation of state:
$p_\mathrm{pt} = 1.01221731\times 10^{35}$, $\epsilon_\mathrm{pt} =
1.01995577\times 10^{15}$ and $\Upsilon_\mathrm{pt} = 0.586578885$.
Two zone representations have been computed for this study with
$N^{<}_\mathrm{parms} =N^{>}_\mathrm{parms}=1$, 2, 3, and 4, thus
resulting in representations with $N_\mathrm{parms}=5$, 7, 9, and 11.
Figure~\ref{f:TwoZoneFits} illustrates the relationships between the
accuracies of the one and two zone spectral and piecewise analytic
representations.  This figure shows that the one zone representations
are more accurate than the two zone representations for the range of
models studied here: $N_\mathrm{parms}\leq 11$.  This figure also
shows that the spectral representations are generally more accurate
than the piecewise analytic representations having the same number of
parameters.
%

%%%%%%%%%%%%%%%%%%%%%%%%%%%%%%%%%%%%%%%%%%%%%%%%%%%%%%%%%%%%%%%%%%%%%%%%%%%%%%%
% Discussion
%%%%%%%%%%%%%%%%%%%%%%%%%%%%%%%%%%%%%%%%%%%%%%%%%%%%%%%%%%%%%%%%%%%%%%%%%%%%%%%
\section{Discussion}
\label{s:Discussion}

In summary, tests have been performed in this study to determine the
accuracy and convergence rates of the spectral and the piecewise
analytic representations of neutron-star equations of state with
strong hadron-quark phase transitions.  These results show that both
types of representation are generally convergent as the number of
parameters is increased.  These representations have comparable
accuracies, with the spectral representations generally being slightly
more accurate than the piecewise analytic representations having the
same number of adjustable parameters.  The two zone spectral
representations should converge exponentially, and should therefore
become more accurate than their one zone counterparts in the limit of
large parameter numbers.  However the tests performed in this study
using eleven or fewer parameters show that the one zone spectral and
piecewise analytic representations are more accurate than their two
zone counterparts with the same number of parameters.  The analysis in
Ref.~\cite{Lindblom2025} shows that very high accuracy observations of
neutron-star masses and radii are needed before equation of state
representations with large numbers of parameters will be useful in
determining the equation of state more accurately.  Therefore it is
not likely that two zone neutron-star representations will be useful
for observational data analysis in the near future.

\begin{figure}[!t]
    \includegraphics[width=0.4\textwidth]
%                    {../PFits/Convergence_Fits.eps}
                    {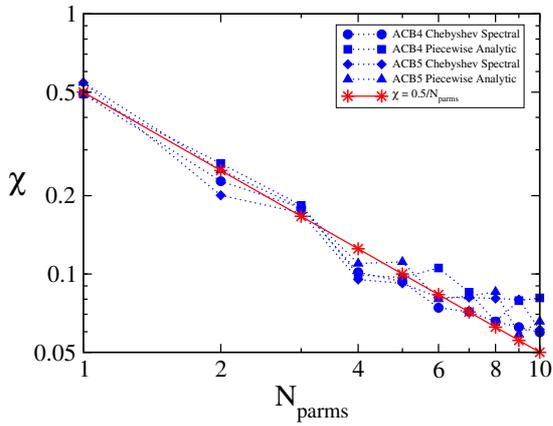}
          \caption{This figure illustrates the convergence rates of
            the representations of the discontinuous ACB4 and ACB5
            equations of state with $\Delta_\mathrm{P}=0$.  The (blue)
            dotted curves reproduce the data shown in
            Fig.~\ref{f:ComparisonFits}, while the (red) solid curve
            is the function $\chi=0.5/N_\mathrm{parms}$.  This
            illustrates that the accuracies of both spectral and the
            piecewise analytic representations converge roughly as
            $1/N_\mathrm{parms}$.
          \label{f:ConvergenceFits}}
\end{figure}
Figure~\ref{f:ComparisonFits} shows that the causal Chebyshev spectral
and the causal piecewise analytic representations of the discontinuous
ACB4 and ACB5 equations of state with $\Delta_\mathrm{P}=0$ have
comparable accuracies.  It is possible that other more accurate
piecewise analytic representations could be found.  However, it is
unlikely that such representations will be significantly more accurate
than the representations studied here.  Figure~\ref{f:ConvergenceFits}
shows that both the spectral and the piecewise analytic
representations used here converge roughly as $1/N_\mathrm{parms}$.
This $1/N_\mathrm{parms}$ convergence rate is the best that can be
expected for piecewise analytic representations of discontinuous
equations of state without any {\it a prior} knowledge of the
locations or the sizes of the discontinuities.  Given that the
representations studied here already converge roughly at this rate, it
is very unlikely that significantly better representations could be
found.

The efficacy of using spectral representations of the neutron-star
equation of state to solve the relativistic inverse stellar structure
problem has been demonstrated in Refs.\cite{Lindblom2012,
  Lindblom2014, Lindblom2018a, Lindblom2025}.  Those studies, however,
could be extended and improved in a number of ways.  The mock
mass-radius data used in those studies were based on equations of
state without the very large hadron-quark type phase transitions
included in this study. Future studies could explore such cases.  The
study in Ref.~\cite{Lindblom2025} used mock noisy mass-radius data
having uniformly distributed observational errors.  Future studies
could improve this by using more realistic observational
errors. 

%%%%%%%%%%%%%%%%%%%%%%%%%%%%%%%%%%%%%%%%%%%%%%%%%%%%%%%%%%%%%%%%%%%%%%%%%%%%%%%
%%%%%%%%%%%%%%%%%%%%%%%%%%%%%%%%%%%%%%%%%%%%%%%%%%%%%%%%%%%%%%%%%%%%%%%%%%%%%%%

\acknowledgments

L.L. was supported in part by the National Science Foundation grants
2012857 and 2407545 to the University of California at San Diego, USA.

%%%%%%%%%%%%%%%%%%%%%%%%%%%%%%%%%%%%%%%%%%%%%%%%%%%%%%%%%%%%%%%%%%%%%%%%%%%%%%%
% References
%%%%%%%%%%%%%%%%%%%%%%%%%%%%%%%%%%%%%%%%%%%%%%%%%%%%%%%%%%%%%%%%%%%%%%%%%%%%%%%
\bibstyle{prd}
\bibliography{../References/References}

\begin{thebibliography}{35}
\expandafter\ifx\csname natexlab\endcsname\relax\def\natexlab#1{#1}\fi
\expandafter\ifx\csname bibnamefont\endcsname\relax
  \def\bibnamefont#1{#1}\fi
\expandafter\ifx\csname bibfnamefont\endcsname\relax
  \def\bibfnamefont#1{#1}\fi
\expandafter\ifx\csname citenamefont\endcsname\relax
  \def\citenamefont#1{#1}\fi
\expandafter\ifx\csname url\endcsname\relax
  \def\url#1{\texttt{#1}}\fi
\expandafter\ifx\csname urlprefix\endcsname\relax\def\urlprefix{URL }\fi
\providecommand{\bibinfo}[2]{#2}
\providecommand{\eprint}[2][]{\url{#2}}

\bibitem[{\citenamefont{Lindblom and Indik}(2012)}]{Lindblom2012}
\bibinfo{author}{\bibfnamefont{L.}~\bibnamefont{Lindblom}} \bibnamefont{and}
  \bibinfo{author}{\bibfnamefont{N.~M.} \bibnamefont{Indik}},
  \bibinfo{journal}{Phys.\ Rev.\ D} \textbf{\bibinfo{volume}{86}},
  \bibinfo{pages}{084003} (\bibinfo{year}{2012}).

\bibitem[{\citenamefont{Lindblom and Indik}(2014)}]{Lindblom2014}
\bibinfo{author}{\bibfnamefont{L.}~\bibnamefont{Lindblom}} \bibnamefont{and}
  \bibinfo{author}{\bibfnamefont{N.~M.} \bibnamefont{Indik}},
  \bibinfo{journal}{Phys.\ Rev.\ D} \textbf{\bibinfo{volume}{89}},
  \bibinfo{pages}{064003} (\bibinfo{year}{2014}).

\bibitem[{\citenamefont{Lindblom}(2014)}]{Lindblom2014a}
\bibinfo{author}{\bibfnamefont{L.}~\bibnamefont{Lindblom}},
  \bibinfo{journal}{AIP Conference Proceedings}
  \textbf{\bibinfo{volume}{1577}}, \bibinfo{pages}{153} (\bibinfo{year}{2014}).

\bibitem[{\citenamefont{Lindblom}(2018{\natexlab{a}})}]{Lindblom2018a}
\bibinfo{author}{\bibfnamefont{L.}~\bibnamefont{Lindblom}},
  \bibinfo{journal}{Phys.\ Rev.\ D} \textbf{\bibinfo{volume}{98}},
  \bibinfo{pages}{043012} (\bibinfo{year}{2018}{\natexlab{a}}).

\bibitem[{\citenamefont{\"Ozel and Freire}(2016)}]{Ozel2016a}
\bibinfo{author}{\bibfnamefont{F.}~\bibnamefont{\"Ozel}} \bibnamefont{and}
  \bibinfo{author}{\bibfnamefont{P.}~\bibnamefont{Freire}},
  \bibinfo{journal}{Ann. Rev. Astron. and Astroph.}
  \textbf{\bibinfo{volume}{54}}, \bibinfo{pages}{401} (\bibinfo{year}{2016}).

\bibitem[{\citenamefont{{T. E. Riley et al.}}(2019)}]{Riley2019}
\bibinfo{author}{\bibnamefont{{T. E. Riley et al.}}},
  \bibinfo{journal}{Astrophys.\ J.\ Lett.} \textbf{\bibinfo{volume}{887}},
  \bibinfo{pages}{L21} (\bibinfo{year}{2019}).

\bibitem[{\citenamefont{{M. C. Miller et al.}}(2019)}]{Miller2019}
\bibinfo{author}{\bibnamefont{{M. C. Miller et al.}}},
  \bibinfo{journal}{Astrophys.\ J.\ Lett.} \textbf{\bibinfo{volume}{887}},
  \bibinfo{pages}{L24} (\bibinfo{year}{2019}).

\bibitem[{\citenamefont{{M. C. Miller et al.}}(2021)}]{Miller2021}
\bibinfo{author}{\bibnamefont{{M. C. Miller et al.}}},
  \bibinfo{journal}{Astrophys.\ J.\ Lett.} \textbf{\bibinfo{volume}{918}},
  \bibinfo{pages}{L28} (\bibinfo{year}{2021}).

\bibitem[{\citenamefont{Biswas}(2021)}]{Biswas2021}
\bibinfo{author}{\bibfnamefont{B.}~\bibnamefont{Biswas}},
  \bibinfo{journal}{Astrophys.\ J.} \textbf{\bibinfo{volume}{921}},
  \bibinfo{pages}{63} (\bibinfo{year}{2021}).

\bibitem[{\citenamefont{{S. Vinciguerra et al.}}(2024)}]{Vinciguerra2024}
\bibinfo{author}{\bibnamefont{{S. Vinciguerra et al.}}},
  \bibinfo{journal}{Astrophys.\ J.} \textbf{\bibinfo{volume}{961}},
  \bibinfo{pages}{62} (\bibinfo{year}{2024}).

\bibitem[{\citenamefont{{T. Salmi et al.}}(2024)}]{Salmi2024}
\bibinfo{author}{\bibnamefont{{T. Salmi et al.}}} (\bibinfo{year}{2024}),
  \bibinfo{note}{arXiv:2406.14466}.

\bibitem[{\citenamefont{{T. Choudhury et al.}}(2024)}]{Choudhury2024}
\bibinfo{author}{\bibnamefont{{T. Choudhury et al.}}},
  \bibinfo{journal}{Astrophys.\ J.\ Lett.}  (\bibinfo{year}{2024}),
  \bibinfo{note}{arXiv:2407.06789}.

\bibitem[{\citenamefont{{K.Chatziioannou et al.}}(2024)}]{Chatziioannou2024}
\bibinfo{author}{\bibnamefont{{K.Chatziioannou et al.}}}
  (\bibinfo{year}{2024}), \bibinfo{note}{arXiv:2407.11153}.

\bibitem[{\citenamefont{De et~al.}(2018{\natexlab{a}})\citenamefont{De,
  Finstad, Lattimer, Brown, Berger, and Biwer}}]{De2018}
\bibinfo{author}{\bibfnamefont{S.}~\bibnamefont{De}},
  \bibinfo{author}{\bibfnamefont{D.}~\bibnamefont{Finstad}},
  \bibinfo{author}{\bibfnamefont{J.~M.} \bibnamefont{Lattimer}},
  \bibinfo{author}{\bibfnamefont{D.~A.} \bibnamefont{Brown}},
  \bibinfo{author}{\bibfnamefont{E.}~\bibnamefont{Berger}}, \bibnamefont{and}
  \bibinfo{author}{\bibfnamefont{C.~M.} \bibnamefont{Biwer}},
  \bibinfo{journal}{Phys.\ Rev.\ Lett.} \textbf{\bibinfo{volume}{121}},
  \bibinfo{pages}{091102} (\bibinfo{year}{2018}{\natexlab{a}}).

\bibitem[{\citenamefont{De et~al.}(2018{\natexlab{b}})\citenamefont{De,
  Finstad, Lattimer, Brown, Berger, and Biwer}}]{De2018a}
\bibinfo{author}{\bibfnamefont{S.}~\bibnamefont{De}},
  \bibinfo{author}{\bibfnamefont{D.}~\bibnamefont{Finstad}},
  \bibinfo{author}{\bibfnamefont{J.~M.} \bibnamefont{Lattimer}},
  \bibinfo{author}{\bibfnamefont{D.~A.} \bibnamefont{Brown}},
  \bibinfo{author}{\bibfnamefont{E.}~\bibnamefont{Berger}}, \bibnamefont{and}
  \bibinfo{author}{\bibfnamefont{C.~M.} \bibnamefont{Biwer}},
  \bibinfo{journal}{Phys.\ Rev.\ Lett.} \textbf{\bibinfo{volume}{121}},
  \bibinfo{pages}{259902} (\bibinfo{year}{2018}{\natexlab{b}}).

\bibitem[{\citenamefont{Chatziioannou}(2020)}]{Chatziioannou2020}
\bibinfo{author}{\bibfnamefont{K.}~\bibnamefont{Chatziioannou}},
  \bibinfo{journal}{Gen.\ Relativ.\ Gravit.} \textbf{\bibinfo{volume}{52}},
  \bibinfo{pages}{109} (\bibinfo{year}{2020}).

\bibitem[{\citenamefont{Read et~al.}(2009)\citenamefont{Read, Lackey, Owen, and
  Friedman}}]{Read:2008iy}
\bibinfo{author}{\bibfnamefont{J.~S.} \bibnamefont{Read}},
  \bibinfo{author}{\bibfnamefont{B.~D.} \bibnamefont{Lackey}},
  \bibinfo{author}{\bibfnamefont{B.~J.} \bibnamefont{Owen}}, \bibnamefont{and}
  \bibinfo{author}{\bibfnamefont{J.~L.} \bibnamefont{Friedman}},
  \bibinfo{journal}{Phys. Rev.} \textbf{\bibinfo{volume}{D79}},
  \bibinfo{pages}{124032} (\bibinfo{year}{2009}).

\bibitem[{\citenamefont{Lindblom}(2010)}]{Lindblom2010}
\bibinfo{author}{\bibfnamefont{L.}~\bibnamefont{Lindblom}},
  \bibinfo{journal}{Phys.\ Rev.\ D} \textbf{\bibinfo{volume}{82}},
  \bibinfo{pages}{103011} (\bibinfo{year}{2010}).

\bibitem[{\citenamefont{Lindblom}(2018{\natexlab{b}})}]{Lindblom2018}
\bibinfo{author}{\bibfnamefont{L.}~\bibnamefont{Lindblom}},
  \bibinfo{journal}{Phys.\ Rev.\ D} \textbf{\bibinfo{volume}{97}},
  \bibinfo{pages}{123019} (\bibinfo{year}{2018}{\natexlab{b}}).

\bibitem[{\citenamefont{Annala et~al.}(2020)\citenamefont{Annala, Gorda,
  Kurkela, Nattila, and Vourinen}}]{Annala2020}
\bibinfo{author}{\bibfnamefont{E.}~\bibnamefont{Annala}},
  \bibinfo{author}{\bibfnamefont{T.}~\bibnamefont{Gorda}},
  \bibinfo{author}{\bibfnamefont{A.}~\bibnamefont{Kurkela}},
  \bibinfo{author}{\bibfnamefont{J.}~\bibnamefont{Nattila}}, \bibnamefont{and}
  \bibinfo{author}{\bibfnamefont{A.}~\bibnamefont{Vourinen}},
  \bibinfo{journal}{Nature Physics} \textbf{\bibinfo{volume}{16}},
  \bibinfo{pages}{907} (\bibinfo{year}{2020}).

\bibitem[{\citenamefont{Lindblom}(2022)}]{Lindblom2022}
\bibinfo{author}{\bibfnamefont{L.}~\bibnamefont{Lindblom}},
  \bibinfo{journal}{Phys.\ Rev.\ D} \textbf{\bibinfo{volume}{105}},
  \bibinfo{pages}{063031} (\bibinfo{year}{2022}).

\bibitem[{\citenamefont{Lindblom and Zhou}(2024)}]{Lindblom2024b}
\bibinfo{author}{\bibfnamefont{L.}~\bibnamefont{Lindblom}} \bibnamefont{and}
  \bibinfo{author}{\bibfnamefont{T.}~\bibnamefont{Zhou}},
  \bibinfo{journal}{Phys.\ Rev.\ D} \textbf{\bibinfo{volume}{110}},
  \bibinfo{pages}{083030} (\bibinfo{year}{2024}).

\bibitem[{\citenamefont{Lindblom}(2024)}]{Lindblom2024}
\bibinfo{author}{\bibfnamefont{L.}~\bibnamefont{Lindblom}},
  \bibinfo{journal}{Phys.\ Rev.\ D} \textbf{\bibinfo{volume}{110}},
  \bibinfo{pages}{043018} (\bibinfo{year}{2024}).

\bibitem[{\citenamefont{Lindblom and Zhou}(2025)}]{Lindblom2025}
\bibinfo{author}{\bibfnamefont{L.}~\bibnamefont{Lindblom}} \bibnamefont{and}
  \bibinfo{author}{\bibfnamefont{T.}~\bibnamefont{Zhou}},
  \bibinfo{journal}{Phys.\ Rev.\ D} \textbf{\bibinfo{volume}{111}},
  \bibinfo{pages}{063024} (\bibinfo{year}{2025}).

\bibitem[{\citenamefont{Boyd}(1999)}]{Boyd1999}
\bibinfo{author}{\bibfnamefont{J.~P.} \bibnamefont{Boyd}},
  \emph{\bibinfo{title}{Chebyshev and {F}ourier Spectral Methods}}
  (\bibinfo{publisher}{Dover Publications}, \bibinfo{year}{1999}),
  \bibinfo{edition}{2nd} ed.

\bibitem[{\citenamefont{Lindblom}(1998)}]{Lindblom98a}
\bibinfo{author}{\bibfnamefont{L.}~\bibnamefont{Lindblom}},
  \bibinfo{journal}{Phys.\ Rev.\ D} \textbf{\bibinfo{volume}{58}},
  \bibinfo{pages}{024008} (\bibinfo{year}{1998}).

\bibitem[{\citenamefont{Paschalidis et~al.}(2018)\citenamefont{Paschalidis,
  Yagi, Alvarez-Castillo, Blaschke, and Sedrakian}}]{Paschalidis2018}
\bibinfo{author}{\bibfnamefont{V.}~\bibnamefont{Paschalidis}},
  \bibinfo{author}{\bibfnamefont{K.}~\bibnamefont{Yagi}},
  \bibinfo{author}{\bibfnamefont{D.}~\bibnamefont{Alvarez-Castillo}},
  \bibinfo{author}{\bibfnamefont{D.~B.} \bibnamefont{Blaschke}},
  \bibnamefont{and}
  \bibinfo{author}{\bibfnamefont{A.}~\bibnamefont{Sedrakian}},
  \bibinfo{journal}{Phys.\ Rev.\ D} \textbf{\bibinfo{volume}{97}},
  \bibinfo{pages}{084038} (\bibinfo{year}{2018}).

\bibitem[{\citenamefont{Reed~Essick and Holz}(2020)}]{Essick2020}
\bibinfo{author}{\bibfnamefont{P.~L.} \bibnamefont{Reed~Essick}}
  \bibnamefont{and} \bibinfo{author}{\bibfnamefont{D.~E.} \bibnamefont{Holz}},
  \bibinfo{journal}{Phys.\ Rev.\ D} \textbf{\bibinfo{volume}{101}},
  \bibinfo{pages}{063007} (\bibinfo{year}{2020}).

\bibitem[{\citenamefont{Blaschke et~al.}(2020)\citenamefont{Blaschke,
  Alvarez-Castillo, Ayriyan, Grigorian, Largani, and Weber}}]{Blaschke2020}
\bibinfo{author}{\bibfnamefont{D.}~\bibnamefont{Blaschke}},
  \bibinfo{author}{\bibfnamefont{D.~E.} \bibnamefont{Alvarez-Castillo}},
  \bibinfo{author}{\bibfnamefont{A.}~\bibnamefont{Ayriyan}},
  \bibinfo{author}{\bibfnamefont{H.}~\bibnamefont{Grigorian}},
  \bibinfo{author}{\bibfnamefont{N.~K.} \bibnamefont{Largani}},
  \bibnamefont{and} \bibinfo{author}{\bibfnamefont{F.}~\bibnamefont{Weber}}, in
  \emph{\bibinfo{booktitle}{Topics on Strong Gravity: A Modern View on Theories
  and Experiments}}, edited by \bibinfo{editor}{\bibfnamefont{C.~A.~Z.}
  \bibnamefont{Vasconcellos}} (\bibinfo{publisher}{World Scientific},
  \bibinfo{year}{2020}), chap.~\bibinfo{chapter}{7}, pp.
  \bibinfo{pages}{207--256}.

\bibitem[{\citenamefont{Glendenning}(1992)}]{Glendenning1992}
\bibinfo{author}{\bibfnamefont{N.~K.} \bibnamefont{Glendenning}},
  \bibinfo{journal}{Phys.\ Rev.\ D} \textbf{\bibinfo{volume}{46}},
  \bibinfo{pages}{1274} (\bibinfo{year}{1992}).

\bibitem[{\citenamefont{Glendenning}(2001)}]{Glendenning2001}
\bibinfo{author}{\bibfnamefont{N.~K.} \bibnamefont{Glendenning}},
  \bibinfo{journal}{Physics Reports} \textbf{\bibinfo{volume}{342}},
  \bibinfo{pages}{392} (\bibinfo{year}{2001}).

\bibitem[{\citenamefont{Abgaryan et~al.}(2018)\citenamefont{Abgaryan,
  Alvarez-Castillo, Ayriyan, Blaschke, and Grigorian}}]{Abgaryan2018}
\bibinfo{author}{\bibfnamefont{V.}~\bibnamefont{Abgaryan}},
  \bibinfo{author}{\bibfnamefont{D.}~\bibnamefont{Alvarez-Castillo}},
  \bibinfo{author}{\bibfnamefont{A.}~\bibnamefont{Ayriyan}},
  \bibinfo{author}{\bibfnamefont{D.}~\bibnamefont{Blaschke}}, \bibnamefont{and}
  \bibinfo{author}{\bibfnamefont{H.}~\bibnamefont{Grigorian}},
  \bibinfo{journal}{Universe} \textbf{\bibinfo{volume}{4}}, \bibinfo{pages}{94}
  (\bibinfo{year}{2018}).

\bibitem[{\citenamefont{Oppenheimer and Volkoff}(1939)}]{Oppenheimer1939}
\bibinfo{author}{\bibfnamefont{J.~R.} \bibnamefont{Oppenheimer}}
  \bibnamefont{and} \bibinfo{author}{\bibfnamefont{G.~M.}
  \bibnamefont{Volkoff}}, \bibinfo{journal}{Phys. Rev.}
  \textbf{\bibinfo{volume}{55}}, \bibinfo{pages}{374} (\bibinfo{year}{1939}).

\bibitem[{\citenamefont{Landau and Lifshitz}(1959)}]{Landau1959}
\bibinfo{author}{\bibfnamefont{L.~D.} \bibnamefont{Landau}} \bibnamefont{and}
  \bibinfo{author}{\bibfnamefont{E.~M.} \bibnamefont{Lifshitz}},
  \emph{\bibinfo{title}{Fluid Mechanics}} (\bibinfo{publisher}{Pergamon Press},
  \bibinfo{year}{1959}).

\bibitem[{\citenamefont{Press et~al.}(1992)\citenamefont{Press, Teukolsky,
  Vetterling, and Flannery}}]{numrec_f}
\bibinfo{author}{\bibfnamefont{W.~H.} \bibnamefont{Press}},
  \bibinfo{author}{\bibfnamefont{S.~A.} \bibnamefont{Teukolsky}},
  \bibinfo{author}{\bibfnamefont{W.~T.} \bibnamefont{Vetterling}},
  \bibnamefont{and} \bibinfo{author}{\bibfnamefont{B.~P.}
  \bibnamefont{Flannery}}, \emph{\bibinfo{title}{Numerical Recipes in
  {FORTRAN}}} (\bibinfo{publisher}{Cambridge University Press},
  \bibinfo{address}{Cambridge, England}, \bibinfo{year}{1992}),
  \bibinfo{edition}{2nd} ed.

\end{thebibliography}

%%%%%%%%%%%%%%%%%%%%%%%%%%%%%%%%%%%%%%%%%%%%%%%%%%%%%%%%%%%%%%%%%%%%%%%%%%%%%%%
\end{document}